\documentclass[12pt]{iopart}
\usepackage{iopams}  
\begin{document}

\title[Cracking and Stability of Non-Rotating Relativistic Spheres with Anisotropic Internal Stresses]{Cracking and Stability of Non-Rotating Relativistic Spheres with Anisotropic Internal Stresses}

\author{B. S. Ratanpal}

\address{Department of Applied Mathematics, Faculty of Technology \& Engineering, The Maharaja Sayajirao University of Baroda, Vadodara - 390001, Gujarat, India}
\ead{bharatratanpal@gmail.com}
\vspace{10pt}

\begin{abstract}
The stability of static uncharged spheres with anisotropic internal stresses is studied in general relativity. It has been noticed that pressure anisotropy plays an important role for stability of stellar structure. It is shown that radial profile of the stress anisotropy can be considered to decide potentially stable/unstable regions. Finally, pontentially stable/unstable regions of two known models of relativistic star have been examined.
\end{abstract}
%
\vspace{2pc}

\section{Introduction}
\noindent In this work, the material composition of the non-rotating relativistic stellar structure is represented by anisotropic term in the stress-energy tensor. The impact of anisotropic stresses on equilibrium configuration of relativistic stellar configuration can be found in poinering work of Bower and Liang\cite{BL} \& Herrera and Santos\cite{HS}. The local anisotropy in the interior of extremely dense object may occur due to existence of type 3A super fluid\cite{R}\cite{KW}, phase transition\cite{S} or presence of electromagnatic fluid\cite{I} and pion condensation\cite{S1}.\\

\noindent The stability of a stellar structure is an important problem in relativistic astrophysics. Chandrasekhar\cite{Chandra1} studied radial perturbation for isotropic fluid spheres in general relativity and derived the pulsation equation to check the stability of fluid spheres. This method was generalized by Dev and Gleiser\cite{DG03} for anisotropic matter distributions. Herrera\cite{Herrera92} and DiPrisco \textit{et. al.}\cite{DFHV94} introduced concept of cracking (or overturning) which describes an alternative way ($-1\leq V^{2}_{s\perp}-V^{2}_{sr}\leq 1$, where $V_{s\perp}$ and $V_{sr}$ are tangential and radial speed of sound respectively) to determine the stability of anisotropic matter distribution in general relativity. In the stellar structure, if no cracking (or overturning) occurs then such structure is identified as potentially stable, since other kind of perturbation may lead to expansion or collapse of stellar configuration (not absolutely stable). Chan {\em et. al.}\cite{C} found that perturbation of density alone does not take anisotropic configuration out of equilibrium but perturbation of both density and local anisotropy can take the system away from equilibrium. Abreu \textit{et. al.}\cite{AHL07} and Gonz$\acute{a}$lez \textit{et. al.}\cite{GNN15} proved that the regions for which $-1\leq V^{2}_{s\perp}-V^{2}_{sr}\leq 0$ are potentially stable and the regions for which $0< V^{2}_{s\perp}-V^{2}_{sr}\leq 1$ are potentially unstable, where $V^{2}_{s\perp}=\frac{dp_{\perp}}{d\rho}$ and $V^{2}_{sr}=\frac{dp_{r}}{d\rho}$.\\

\noindent In this work, the role of anisotropy in describing potentially stable/unstable regions has been studied. With simple calculations, it has been shown that potentially stable/unstable regions can be recognized from gradient of anisotropy with respect to radial variable $r$.
\section{Static, Uncharged, Anisotropic Matter Distribution}
\label{sec:2}
\noindent The interior spacetime metric for static spherically symmetric stellar structure is considered as 
\begin{equation}\label{InMetric}
ds^{2} = e^{\nu(r)}dt^{2}-e^{\lambda(r)}dr^2-r^2\left(d\theta^2+\sin^2\theta d\phi^2\right).
\end{equation}
We write the energy momentum tensor for anisotropic matter distribution in the form
\begin{equation}\label{EMTensor}
T_{ij} = diag\left(\rho,\;-p_{r},\;-p_{\perp},\;-p_{\perp}\right),
\end{equation}
where $\rho$ represents energy-density, $p_{r}$ and $p_{\perp}$ represent radial and tangtial pressures respectively. The magnitude of anisotropy is defined as
\begin{equation}\label{S}
\Delta = p_{\perp}-p_{r}.
\end{equation}

\noindent For a physically plausible matter configuration radial and tangential speed of sound should be positive and less than 1, i.e. $0\leq V^2_{sr}\leq 1$ and $0\leq V^2_{s\perp}\leq 1$ respectively. If $V^2_{s\perp}=0$, $V^2_{sr}=1$ then $V^2_{s\perp}-V^2_{sr}=-1$ and if $V^2_{s\perp}=1$, $V^2_{sr}=0$ then $V^2_{s\perp}-V^2_{sr}=1$ hence,
\begin{equation}\label{SV}
	Min\left(V^2_{s\perp}-V^2_{sr}\right)=-1,\;\;\;\;\;Max\left(V^2_{s\perp}-V^2_{sr}\right)=1.
\end{equation}
\textbf{Theorem:} For a static spherically symmetric interior spacetime metric (\ref{InMetric}) with energy-momentum tensor (\ref{EMTensor}) with decreasing density with respect to radial variable $r$, the potentially stable regions are those for which gradient of anisotropy with respect to radial variable $r$ is greater than or equal to zero and potentially unstable regions are those for which gradient of anisotropy with respect to radial variable $r$ is negative.\\\\
\textbf{Proof:} Assume that the gradient of anisotropy with respect to radial variable $r$ is greater than or equal to zero. This implies that $\frac{d\Delta}{dr}\geq 0$. Since $\frac{d\rho}{dr}\leq 0$, we get $\frac{d\Delta}{dr}/\frac{d\rho}{dr}\leq 0$, implying that $\frac{d\Delta}{d\rho}\leq 0$. By definition of anisotropy, we get $\frac{d}{d\rho}\left(p_{\perp}-p_{r} \right)\leq 0$ leading to the inequality
\begin{equation}\label{Ineq1}
	V^2_{s\perp}-V^2_{sr} \leq 0.
\end{equation}
Hence from (\ref{SV}) and (\ref{Ineq1}) $-1\leq V^2_{s\perp}-V^2_{sr}\leq 0$, which represents potentially stable regions as described by Abreu \textit{et. al.}\cite{AHL07}.\\
In the similar manner it can be proved that potentially unstable regions are those for which gradient of anisotropy with respect to radial variable $r$ is negative.
\section{Analysis of Two Known Models}
\label{sec:3}
\noindent To check the stability of non-rotating anisotropic matter distribution many authors (\cite{R1},  \cite{DRB}, \cite{GPPU}) have used condition prescribed by Abreu \textit{et. al.}. The alternative way to check this condition is described in theorem in previous section which says if gradient of anisotropy with respect to radial variable $r$ is greater than or equal to 0 then $-1\leq V^2_{s\perp}-V^2_{sr}\leq 0$ leading to the stellar configuration to be potentially stable and if gradient of anisotropy with respect to radial variable $r$ is negative then $0< V^{2}_{s\perp}-V^{2}_{sr}\leq 1$ leading to stellar configuration to be potentially unstable. We now apply this theorem to two known models.\\

\noindent 1. Pant \textit{et. al.} \cite{PGBP} model: In their work anisotropy $\left(\Delta=p_{\perp}-p_{r}\right)$ (Fig. 9 in \cite{PGBP}) is increasing in radially outward direction. Hence their model is potentially stable as per theorem decribed in section 2.\\

\noindent 2. Malaver \textit{et. al.} \cite{MEG} model: In their work also anisotropy $\left(\Delta=p_{\perp}-p_{r}\right)$ (Fig. 5 in \cite{MEG}) is increasing for $\omega=-\frac{7}{10}$, $\omega=-\frac{1}{2}$ while it is decreasing for $\omega=-\frac{2}{5}$, $\omega=-\frac{1}{3}$ and $\omega=-\frac{1}{4}$. Hence their model is potentially stable for $\omega=-\frac{7}{10}$ and $\omega=-\frac{1}{2}$, while is model potentially unstable for $\omega=-\frac{2}{5}$, $\omega=-\frac{1}{3}$ and $\omega=-\frac{1}{4}$ as per theorem described in section 2.

\section{Conclusion}
\label{sec:4}
The results presented in this work suggest that by looking at the graph of anisotropy potentially stable/unstable regions can be identified. However it should be noted that absolute stable regions can not be identified from nature of anisotropy. Hence further analysis is required connecting work of Chardrasekhar \cite{Chandra1}\cite{Chandra2}\cite{Chandra3} and Herrera's cracking method \cite{Herrera92}.
\section*{Acknowlegement}
BSR would like to thank IUCAA, Pune for the facilities and hospitality provided to them where the part of work was carried out. BSR is also thankful to referee and V. O. Thomas for useful suggestions.
\section*{Data Availability}
No new data were created or analysed in this study.

\end{document}